\documentstyle[12pt]{article}
\newcommand{\bce}{\begin{center}}
\newcommand{\ece}{\end{center}}
\newcommand{\beq}{\begin{equation}}
\newcommand{\eeq}{\end{equation}}
\newcommand{\bea}{\vspace{0.25cm}\begin{eqnarray}}
\newcommand{\eea}{\end{eqnarray}}

\newcommand{\ba}{\begin{array}}
\newcommand{\ea}{\end{array}}

\newcommand{\doublespace}{
    \renewcommand{\baselinestretch}{1.6}\large\normalsize}

\def\lsim{\mathrel{\rlap{\lower4pt\hbox{\hskip1pt$\sim$}}
    \raise1pt\hbox{$<$}}}         %less than or approx. symbol
\def\gsim{\mathrel{\rlap{\lower4pt\hbox{\hskip1pt$\sim$}}
    \raise1pt\hbox{$>$}}}         %greater than or approx. symbol

\def\beq{\begin{equation}}
\def\endeq{\end{equation}}
\def\bea{\begin{eqnarray}}
\def\arr{\begin{eqnarray}}
\def\endarr{\end{eqnarray}}

\makeindex
\begin{document}
%\large
\phantom{.}\hspace{11.0cm} SISSA \\
\phantom{.}\hspace{11.0cm} May, 1995
\vspace{2cm}
\begin{center}
{\bf \LARGE The longitudinal asymmetry of the $(e,e'p)$ missing
momentum distribution as a signal of
color transparency
\\}
\vspace{1.5cm}
{\bf \large O.Benhar$^{1)}$,
 S.Fantoni$^{2,3)}$,
N.N.Nikolaev$^{4,5)}$,
J.Speth$^{4)}$, A.A.Usmani$^{2)}$,
B.G.Zakharov$^{5)}$ } \bigskip\\
{\small \sl
$^{1)}$INFN, Sezione Sanit\`{a}, Physics Laboratory,
 Istituto Superiore di Sanit\`{a}. I-00161 Roma, Italy \\
$^{2)}$Interdisciplinary Laboratory, SISSA, INFN,
Sezione di Trieste. I-34014, Trieste, Italy \\
$^{3)}$International Centre for Theoretical Physics,
Strada Costiera 11, I-34014, Trieste, Italy\\
$^{4)}$IKP(Theorie), Forschungszentrum  J\"ulich GmbH.\\
D-52425 J\"ulich, Germany \\
$^{5)}$L.D.Landau Institute for Theoretical Physics. \\
GSP-1, 117940, ul.Kosygina 2. V-334 Moscow, Russia
%-----------------------------------------------------------
\vspace{1cm}\\}
{\bf \large Abstract \bigskip\\}
\end{center}
We use multiple scattering theory to evaluate
the $Q^{2}$ dependence of the forward-backward asymmetry, $A_{z}$,
 of the missing momentum distribution in quasielstic $A(e,e'p)$
scattering, which is expected to be affected
by color transparency. The novel features of our analysis
are a consistent treatment
of the structure of the ejectile state formed after absorption of
the virtual photon by the struck proton and a careful evaluation
of the background asymmetry induced by the nonvanishing real part
of the proton-nucleon scattering amplitude. We find that
 the absolute magnitude of $A_{z}$ is dominated by this
background at the $Q^{2}$'s attainable at the Continuous Electron Beam
 Accelerator Facility. However, the $Q^{2}$ dependence of the asymmetry
is sensitive to the onset of color transparency, whose observation
 in high statistics experiments appears to be feasible, particularly using
 light targets such as $^{12}C$ or $^{16}O$.

%-----------------------------------------------------------
\newpage
\doublespace

%-----------------------------------------------------------
%\section{Introduction}
%-----------------------------------------------------------

The recent $A(e,e'p)$ experiment carried out at SLAC by the NE18 collaboration
\cite{NE18} did not
observe any substantial $Q^{2}$ dependence of the integrated nuclear
 transparency
 up to $Q^{2} \lsim 7$ GeV$^{2}$.
On the other hand, theory predicts that final state interactions (FSI)
should vanish at asymptotically high $Q^{2}$ on account of color transparency
 (CT) \cite{M1,Brodsky1}.
{}From the point of view of multiple scattering
theory (MST), CT corresponds to a cancellation
between the rescattering amplitudes with elastic (diagonal)
and inelastic (off-diagonal) intermediate states.
The NE18 finding was anticipated in ref.\cite{JTr}, where it was argued
that in the integrated nuclear
transparency the effect of the inelastic rescatterings
of the struck proton is still weak in the
region $Q^{2} \lsim 7$ GeV$^{2}$, covered by NE18 and
achievable by the forthcoming experiments at CEBAF.
The cancellation of the diagonal and off-diagonal rescatterings
depends upon the missing momentum \cite{K1,JAz,B1}, which can be
used to enhance the contribution from inelastic rescatterings
and make the onset of CT observable at CEBAF.
 The enhancement and/or suppression
of the contribution of inelastic rescatterings result in a non-vanishing
forward-backward asymmetry of the missing momentum distribution
, $w(\vec{p})$,
defined as \cite{B1}
\beq
A_{z}(x,y)=\frac{N_{+}-N_{-}}{N_{+}+N_{-}}\,,
\label{eq:1}
\eeq
 where $N_{+}$ $(N_{-})$ is the number of events in the
kinematical region $x<p_{z}<y$ $(-y<p_{z}<-x)$ and the
$z$-axis is parallel to the $(e,e')$ momentum transfer $\vec{q}$.

Unfortunately, besides CT, there is another competing source of
asymmetry associated with FSI: the nonvanishing ratio between
the real and the imaginary
 part of the elastic proton-nucleon ($pN$) scattering amplitude, $\alpha_{pN}$.
 The
forward-backward asymmetry, associated with $\alpha_{pN}$, disregarded
 in refs.\cite{K1,JAz,B1}, was recently found to be sizeable \cite{NSZ}.
 The role played by the real part of the rescattering amplitude in electron
nucleus scattering at high $Q^2$ has been also pointed out in
 ref.\cite{gangofsix},
within the context of an analysis of the inclusive data from SLAC.

In this paper we give a quantitative evaluation
of how much the CT signal can be obscured by the effect
of a nonzero $\alpha_{pN}$. From the technical point of view, one must
calculate $A_{z}$ taking into account the
coupled-channel treatment of the off-diagonal rescatterings and the
nonzero $\alpha_{pN}$ on the same footing. Besides including the effect of
$\alpha_{pN}$, we improve upon the previous works treating the coherency
 properties
of off-diagonal rescatterings in a more accurate manner.

The missing momentum distribution equals
\bea
w(\vec{p})&=&\frac{1}{(2\pi)^{3}}\int
d^{3}\vec{r}_{1}d^{3}\vec{r}_{1}^{\,'}
d^{3}\vec{r}_{2}\cdots d^{3}\vec{r}_{A}
\rho_{p}(
\vec{r}_{1}, \vec{r}_{2}, \cdots , \vec{r}_{A};\vec{r}_{1}^{\,'},
 \vec{r}_{2}, \cdots , \vec{r}_{A})
\nonumber\\
& & \,\,\,\,\,\,\,\,\,\,\,\,\,\,\,\,\,\,\, \times
S^{*}( \vec{r}_{1}^{\,'}, \vec{r}_{2}, \cdots , \vec{r}_{A})
S( \vec{r}_{1}, \vec{r}_{2}, \cdots , \vec{r}_{A})
\exp[i\vec{p}(\vec{r}_{1}^{\,'}-\vec{r}_{1})]\,,
\label{eq:2}
\eea
where
$\rho_{p}(
\vec{r}_{1}, \vec{r}_{2}, \cdots , \vec{r}_{A};\vec{r}_{1}^{\,'},
 \vec{r}_{2}, \cdots , \vec{r}_{A})$
is the proton $A$-body semidiagonal density matrix \cite{feenberg} and
\beq
S( \vec{r}_{1}, \vec{r}_{2}, \cdots , \vec{r}_{A})=
\frac{\langle p|\hat{S}_{3q}(\vec{r}_{1},\vec{r}_{2},...,\vec{r}_{A})
|E\rangle}
{\langle p|E\rangle}\,.
\label{eq:3}
\eeq
In eq.(\ref{eq:3}), $\hat{S}_{3q}(\vec{r}_{1},
\vec{r}_{2},...,\vec{r}_{A})$ is the
operator describing the evolution of the ejectile state
during its propagation through the nuclear medium and the ejectile
state $|E\rangle$ is a three-quark state formed from the struck
proton after absorption of the virtual photon at position
$\vec{r}_{1}$.
 In terms of the electromagnetic current operator
$\hat J_{em}$,
$
|E\rangle=\hat{J}_{em}(Q)|p\rangle=\sum\limits_{i}
|i\rangle\langle i|J_{em}(Q)|p\rangle=\sum\limits_{i}
G_{ip}(Q)|i\rangle\,,
$
where the sum includes the complete set of three-quark states and
$G_{ip}(Q)=\langle i|J_{em}(Q)|p\rangle=\langle i|E\rangle$ denotes both
 the electromagnetic form factor of the proton ($|i\rangle = |p\rangle$) and
 all the transition form factors associated with the electroexcitation
processes $e~+~p~\rightarrow~e'~+~i$
\cite{JAz}.

We will consider the forward-backward asymmetry at
moderate missing momentum $p\lsim 200$ MeV, where
the main effect of short range $NN$ correlations in the
 initial
state is an overall renormalization of the momentum distribution \cite{NNcor},
 which does not affect the calculation of $A_{z}$. Although $NN$
 correlations also contribute to the integrated transparency
\cite{Ben92,NNNcor}, their effect
is expected to be less important in $A_{z}$. Neglecting exchange
and dynamical correlations between the struck nucleon and the
spectator nucleons, the proton $\nu$-body semidiagonal density matrix,
$\rho(
\vec{r}_{1}, \vec{r}_{2}, \cdots , \vec{r}_{\nu};\vec{r}_{1}^{\,'},
 \vec{r}_{2}, \cdots , \vec{r}_{\nu})$
 takes the factorized form
$\rho_{p} (\vec{r}_{1},\vec{r}_{1}^{\,'})\rho (\vec{r}_{2},
 \cdots ,\vec{r}_{\nu})$,
where $\rho (\vec{r}_{2}, \cdots , \vec{r}_{\nu})$ is the $(\nu-1)$-body
 diagonal density matrix which can be recast in the form \cite{schiavilla}
$\prod_{i=2}^{\nu} \rho(\vec{r}_{i})
g(\vec{r}_{2}, \cdots ,\vec{r}_{\nu})$
where $\rho(\vec{r}_{i})=n_{A}(\vec{r}_{i})/A$ is the one-body density and
the $(\nu-1)$-body distribution function \cite{feenberg},
 $g(\vec{r}_{2}, \cdots ,\vec{r}_{\nu})$,
 brings in the effect of $NN$ correlations. We also neglect
$NN$ correlations among the spectator nucleons
in this first study of longitudinal asymmetry. This implies that
$g(\vec{r}_{2}, \cdots ,\vec{r}_{\nu})=1$
 and one ends up with the results of
 ref.~[8] that at
moderate missing momenta $p \lsim 200$ MeV$/$c, the
integrand of (\ref{eq:2}) - the FSI-modified
one-body density matrix - can be approximated as
\beq
\rho_{p}(\vec{r}_{1},\vec{r}_{1}^{\,'})
\int
d^{3}\vec{r}_{2} \cdots d^{3}\vec{r}_{A}
\, \prod_{i=1}^{A}\rho (\vec{r}_{i})
S^{*}( \vec{r}_{1}^{\,'},
 \vec{r}_{2}, \,\, \cdots , \vec{r}_{A})
S( \vec{r}_{1}, \vec{r}_{2}, \cdots , \vec{r}_{A})
=\rho_{p}(\vec{r}_{1},\vec{r}_{1}^{\,'})\Phi(\vec{r}_{1},\vec{r}_{1}^{\,'})
\label{eq:4}
\eeq
where,
$
\rho_{p}(\vec{r}_{1}, \vec{r}_{1}^{\,'})=\frac{1}{Z} \sum_{n}
\phi_{n}^{*}(\vec{r}_{1}^{\,'})\phi_{n} ( \vec{r}_{1})
$
is the proton shell model one-body density matrix and
$\phi_{n}$ are the shell model wavefunctions.

In the MST, the FSI operator $\hat{S}_{3q}^{*}\hat{S}_{3q}$
can be very schematically
represented as $S_{3q}^{*}S_{3q}=\prod_{j=2}^{A}
(1-\hat{\Gamma}^{*}_{1'j})(1-\hat{\Gamma}_{1j})$, where
the profile function
$\hat\Gamma$ is an operator acting on the state describing the internal
structure of the three-quark system [8].
Then, the calculation of $\Phi(\vec{r}_{1},\vec{r}_{1}^{\,'})$
involves a sum of diagrams, as shown in Fig.~1.
Every dotted line attached to the straight-line trajectory originating from
 the point $\vec{r}_{1}$ ($\vec{r}_{1}^{\,'}$)
 denotes a profile function $\hat\Gamma_{1j}$ ($\hat\Gamma^{*}_{1^{'}j}$),
 where the subscripts $1j(1^{'}j)$ stands for the
 position of the $j$-th spectator and its projection on the trajectory.
The interaction between the two trajectories, due to terms
$\hat\Gamma_{1j}\hat\Gamma^{*}_{1^{'}j}$ like the one shown in
Fig.~1b, makes the FSI factor
$\Phi(\vec{r}_{1},\vec{r}_{1}^{\,'})$
 a non factorizable
function of $\vec{r}_{1}$ and $\vec{r}_{1}^{\,'}$.
It can be shown that the non factorizable part of
$\Phi(\vec{r}_{1},\vec{r}_{1}^{\,'})$
is only connected with the incoherent elastic
rescatterings, which come into play only at $p\gsim 200$ MeV \cite{NSZ}.
 For this reason, at
$p\lsim 200$ MeV, we
can neglect $\hat\Gamma_{1j}\hat\Gamma^{*}_{1^{'}j}$ terms
in the FSI operator $\hat{S}^{*}_{3q}
\hat{S}_{3q}$ and obtain  the factorizable $
\Phi(\vec{r}_{1},\vec{r}_{1}^{\,'})=S^{*}_{coh}(\vec{r}_{1})S_{coh}
(\vec{r}_{1}^{\,'})$ and the
following expression for
$w(\vec{p})$,
\beq
w(\vec{p})=
\frac{1}{Z}\sum\limits_{n}
\left|\int d^{3}\vec{r}\phi_{n}(\vec{r})\exp[-i\vec{p}\vec{r}]
S_{coh}(\vec{r})\right|^{2}\,,
\eeq
where
\beq
S_{coh}(\vec{r}_{1})=
\int \prod\limits_{j=2}^{A}\rho(\vec{r}_{j}) d^{3}\vec{r}_{j}
\frac{\langle p|\hat{S}_{3q}(\vec{r}_{1},\vec{r}_{2},...,\vec{r}_{A})
|E\rangle}
{\langle p|E\rangle}\,\,  .
\label{eq:5}
\eeq
Finally, using the Glauber form of $\hat{S}_{3q}$,
the coupled channel MST series for $S_{coh}(\vec{r})$ in terms of the
$\nu$-fold off-diagonal rescatterings can be cast in the form
$
%\beq
S_{coh}(\vec{r})=
\sum\limits_{\nu=0}^{\infty}
S_{coh}^{\nu}(\vec{r})\,,
%\label{eq:9}
%\eeq
$
where the first few terms read
%****************************************************
\bea
S_{coh}(\vec{b},z)&=&1- \exp[i k_{i_{1}p}z]
\sum_{i_{1}}\frac{\sigma^{'}_{pi_{1}}}{2}
 \frac{\langle i_{1}|E\rangle}{\langle p|E\rangle}
\int\limits_{z}^{\infty}dz_{1} n_{A}(\vec{b},z_{1})
\exp[i k_{pi_{1}}z_{1}-{1 \over 2}t(\vec{b},z_{1},z)\sigma_{i_{1}i_{1}}]
\nonumber\\
&+&
\exp[i k_{i_{1}p}z]
\sum_{i_{1}i_{2}}
\frac{\sigma^{'}_{pi_{2}}}{2}
\frac{\sigma^{'}_{i_{2}i_{1}}}{2}
 \frac{\langle i_{1}|E\rangle}{\langle p|E\rangle}
\int\limits_{z}^{\infty}dz_{1} n_{A}(\vec{b},z_{1})
\exp[i k_{i_{2}i_{1}}z_{1}-{1 \over 2}t(\vec{b},z_{1},z)\sigma_{i_{1}i_{1}}]
\nonumber\\
&\times &
\int \limits_{z_{1}}^{\infty}dz_{2} n_{A}(\vec{b},z_{2})
\exp[i k_{pi_{2}}(z_{2}-z_{1})-{1 \over 2}t(\vec{b},z_{2},z_{1})
\sigma_{i_{2}i_{2}}]
\nonumber\\
&+&\cdots  ,
\label{eq:6}
\eea
%*****************************************************
where $\vec{r}=(\vec{b},z)$,
$\sigma^{'}_{ik}=\sigma_{ik}-\delta_{ik}\sigma_{ii}$,
the matrix $\hat{\sigma}=2\int d^{2}\vec{b}\hat\Gamma(\vec{b})$
is connected with the forward diffraction
scattering matrix $\hat{f}=i\hat{\sigma}$ (notice that
Re $\sigma_{ii}=\sigma_{tot}(iN)$)
and
$t(\vec{b},z_{2},z_{1})=\int_{z_{1}}^{z_{2}}dz n_{A}(\vec{b},z)\,$
is the partial optical thickness. Eq.(\ref{eq:6})
shows contributions from up to two off-diagonal rescatterings and any
number of elastic rescatterings in between, higher order off-diagonal
rescatterings are implied by the dots.
The onset of CT is controlled by the
oscillating factors $\exp[i k_{i_{n}i_{n-1}}z_{n}]$, taking into account
the longitudinal momentum transfer associated with each off-diagonal
transition $i_{n-1}N\rightarrow i_{n}N$ \cite{Gribov} according to
\beq
k_{i_{n}i_{n-1}}=\frac{m_{i_{n}}^{2}-m_{i_{n-1}}^2}
{2\varepsilon}\,,
\label{eq:7}
\eeq
where $m_{i}$ denotes the mass of the proton excitation $|i\rangle$ and
$\varepsilon$ is the  energy of the struck proton in the
laboratory frame.

%%%%%%%%%%%%%%%%%%%%%%%%%%%%%%%%%%%%%%%%%%%%%%%%%%%%%%%%%%%%%
%To proceed with evaluation of $w(\vec{p})$
%making use of Eq.~(\ref{eq:6}) we
%need the forward diffraction
%scattering matrix $\hat{f}$ and must choose
%the initial wave function $|E\rangle$.
%If we assume the dominance of the hard mechanism in the
%electromagnetic form factors \cite{Brodsky2}, then it can be shown that
%the corresponding effective ejectile state, which only includes
%the states satisfying the coherency constraint, shall become a
%pointlike state in the limit of $Q^{2}\rightarrow \infty$
%\cite{NNNJETP,NNZCEBAF}. In the subsequent analysis we assume the
%pointlike initial state $|E\rangle$.
%%%%%%%%%%%%%%%%%%%%%%%%%%%%%%%%%%%%%%%%%%%%%%%%%%%%%%%%%%%%%

{}From the point of view of CT, the most
important ingredient in the coupled-channel
formalism is the Pomeron contribution to the
matrix $\hat{\sigma}$. As in ref. \cite{JTr}, we will construct $\hat{\sigma}$
using the oscillator quark-diquark model of the proton.
As a result, the Pomeron contribution to the matrix element $\sigma_{ik}$
can be written as
\beq
{\rm Im}~f(kN\rightarrow iN)=
{\rm Re}~\sigma_{ik}^{P}=\int dz d^{2} \rho \,
\Psi_{i}^{*}(\vec{\rho},z)\sigma(\rho)\Psi_{k}(\vec{\rho},z)\,,
\label{eq:8}
\eeq
where $\Psi_{i,k}$ are the oscillator wave functions describing
the quark-diquark states and $\sigma(\rho)$ is the dipole cross
section describing the interaction of the quark-diquark system with a nucleon,
which was taken in the form
$
\sigma(\rho)=\sigma_{0}\left [
1-\exp\left (-\frac{\rho^{2}}{R_{0}^{2}}\right )\right ]\,.
$
Following ref.\cite{JTr}, we set $\sigma_{0}=2\sigma_{tot}(pN)$
and adjust $R_{0}$ to reproduce $\sigma_{tot}^{exp}(pN)$.
For the diagonal $iN \rightarrow iN$ scattering we take
${\rm Re}~f(iN\rightarrow iN)=
-{\rm Im}~\sigma_{ii}=\frac{1}{2}
\left(\alpha_{pp}\sigma_{tot}(pp)+
\alpha_{pn}\sigma_{tot}(pn)\right)$. Even though this choice can be
justified within the framework of the dual parton model \cite{DTU},
we are fully aware that it should only be regarded as an estimate.
 The real parts
${\rm Re}~f(iN\rightarrow kN)$ of the off-dagonal amplitudes are not
known experimentally. We shall give an estimate of the sensitivity of
our results to these quantities and the associated uncertainty.

For the oscillator frequency of the quark-diquark system we
use the value $\omega_{qD}=0.35$ GeV,
 leading to a realistic mass spectrum of the proton
excitations.
 Since we start with a
 $\sigma(\rho)$ which vanishes as $\rho \rightarrow 0$,
our diffraction matrix satisfies the CT sum rules \cite{JAz,NNZCEBAF}
by construction,
 and leads to vanishing FSI in $(e,e'p)$ at
asymptotically large $Q^{2}$. Moreover, with the above form of
 $\sigma(\rho)$, we obtain a diffraction matrix yielding
 a realistic description of the mass spectrum
 observed in diffractive $pN$ scattering \cite{NNZCEBAF}. For instance,
the present model gives a value of the ratio between the diffractive
and the elastic $pN$ cross sections in agreement with
the experimental data: $\sigma_{diff}(pp)/\sigma_{el}(pp)
\approx 0.25$ \cite{DD}.
The effect of the truncation of the sum over the intermediate states
has been evaluated by direct $z$-integration in eq.~(\ref{eq:6}). This
procedure is more accurate than the effective diffraction matrix approximation
 used in earlier works \cite{JTr,JAz}. The main
difference is that the direct evaluation of the $z$-integrations
in eq.(\ref{eq:6}) produces a somewhat weaker suppression of the
contribution of heavy intermediate states and a somewhat faster
onset of CT effects. Since our diffraction matrix gives a
reasonable description of diffractive $pN$ scattering,
 we believe that the multiple scattering approach employed in this paper
is a good tool for a quantitative study of the onset
of CT in $(e,e'p)$ reactions.
%Evidently, this onset of CT
%is controlled by the contribution
%of inelastic rescatterings of the low-mass excitations of the
%proton, $m^{*^2} - m_{p}^{2} \lsim Q^{2}/R_{A}m_{p}$.

%***************************************************
%{\bf
Besides the matrix $\hat\sigma$, the evaluation of $w(\vec{p})$ requires
 the initial wave function $|E\rangle$.
 Assuming the dominance of the hard mechanism in the
electromagnetic form factors \cite{Brodsky2}, it can be shown that
the projection of the ejectile onto the subspace of hadronic
mass eigenstates which
satisfy the coherency constraint
$m^{*^2} - m_{p}^{2} \lsim Q^{2}/R_{A}m_{p}$,
coincides with the similar projection of
a compact state, with transverse size $\rho\sim 1/Q$
 \cite{NNNJETP,NNZCEBAF}.
For this reason, the latter can be taken for $|E\rangle$, and
in our analysis we use the initial wave
function of the form
$\langle\rho|E\rangle\propto \exp(-C\rho^{2}Q^{2})$,
 with $C=1$. It is worth noting that
the missing momentum distribution is not
sensitive to the specific choice of $C$ as long as
$C\gsim 1/Q^{2}\rho_{o}^{2}\,$,
where $\rho_{o}$ denotes the position of
 the first node in the wave functions of the excited states
satisfying the coherency requirement.
 In the region of $2\lsim Q^{2}\lsim 20$ GeV$^{2}$, that we discuss
 in the present paper, the numerical
results are practically independent of the parameter $C$
for $C\gsim 0.05$.
%}
%***************************************************

Before discussing our numerical results, a comment on
the treatment of $A_{z}$ of refs.\cite{K1,B1,B2} is in order. First,
these authors have not accounted for the large contribution to $A_{z}$
coming from the nonzero $\alpha_{pN}$. Second, we disagree with the
treatment of the initial state $|E\rangle$ suggested in ref.\cite{K1}.
 The formalism of the coupled channel eikonal wave equations
used in \cite{K1,B1,B2} is equivalent to our coupled channel multiple
scattering series of eq.(\ref{eq:6}). However, it has to be noticed that,
within
 the
Plane Wave Impulse Approximation (PWIA), eq.~(\ref{eq:6}) combined with our
 definition
of the ejectile wave function $|E\rangle$ gives
 $d\sigma(A(e,e'i))/d\sigma(A(e,e'p)) =
B_{i}^{2}(\vec{p})/B_{p}^{2}(\vec{p})$, where
\beq
%|\langle i|E(\vec{p})\rangle|^{2}
B_{i}^{2}(\vec{p})\propto \int d^{3}\vec{k}
n(k_{x},k_{y},k_{z})\delta\left(k_{z}-p_{z}-
\frac{m^{2}-m_{i}^{2}}{2\epsilon}\right)G_{ip}^{2}(Q^{2})\,,
\label{eq:9}
\eeq
and $n(\vec{p})$ is the nucleon momentum distribution. Eq.~(\ref{eq:9})
shows the standard kinematical correlation between the mass of the
excited baryon state $|i\rangle$ and the missing momentum, which
can produce either enhancement or suppression of the resonance
abundance in the final state. The procedure used in refs.\cite{K1,B1,B2}
amounts to a redefinition of the ejectile state
\beq
|E(\vec{p})\rangle = \sum_{i}G_{ip}(Q)
{B_{i}(\vec{p})\over B_{p}(\vec{p})}|i\rangle\,,
\label{eq:10}
\endeq
leading to a $\vec{p}$ dependence in ejectile wave
function (\ref{eq:10}) that
 is not born out by the quantum-mechanical treatment of the Fermi
motion. It corresponds to a double counting and evidently does not
give the correct PWIA limit of the production cross sections.
Consequently, the validity of the numerical results of refs.~\cite{K1,B1,B2}
appears to be questionable even within the crude models employed to describe
 the diffraction matrix.

The main results of our work are presented in Fig.~2.
 Following \cite{B1} we  have calculated $A_{z}(x,y)$ assuming
 parallel kinematics, i.e. taking in eq.(\ref{eq:1})
$N_{+}=\int_{x}^{y} dp_{z}w(\vec{p}_{\perp}=0,p_{z})$ and
$N_{-}=\int_{-y}^{-x} dp_{z}w(\vec{p}_{\perp}=0,p_{z})$.
 Fig. 2 shows $A_{z}$ calculated for two
kinematical windows,
 $(x,y)=(0,200)$ and $(x,y)=(50,200)$ MeV, and two different targets,
 $^{16}O$ and $^{40}Ca$.
The parameters of the $pp$- and $pn$-amplitudes were taken
from the recent review of ref.\cite{Lehar}. The number of the included
excited  states and the off-diagonal
rescatterings were equal to 4 and 3 respectively. Contributions
 from higher excitations and rescatterings with $\nu >3$ have been found
 to be negligible at $Q^{2}\lsim 20$ GeV$^{2}$. In the region of
$Q^{2}\lsim 5$ GeV$^{2}$ the effect of CT is almost exhausted
by the first excitation.
To illustrate the contribution of the elasic intermediate state
 to $A_{z}$,
 the results obtained using the Glauber model are also shown
in Fig. 2. In this case the only source of the $z$-asymmetry of
the missing momentum distribution is the nonzero
$\alpha_{pN}$.
As one can see from Fig. 2, in the region of
$Q^{2}\sim 2-5$ GeV$^{2}$, relevant to
 the experimental study of CT effects at CEBAF, the inelastic intermediate
states are only responsible for 25-30 \% (for $^{16}O$)
and 10-15 \% (for $^{40}Ca$)  of $A_{z}$ for both the kinematical
windows in $p_{z}$. The large contribution
 of the elastic rescatterings to $A_{z}$ can make it difficult
 to extract definite conclusions on the onset of CT from measurements of
the magnitude of $A_{z}$ at CEBAF. According to our results,
 the situation is particularly unfavourable in the case of
heavy targets. However, in light nuclei the $Q^{2}$-dependence
of $A_{z}$ is dominated by CT.

Since CT effects on $A_{z}$ are small, assessing  the posssibility
of their observation at CEBAF energies requires a study of the sensitivity
of the theoretical predictions to the unknown ratios
between the real and the imaginary parts of the resonance-nucleon amplitudes.
At $Q^{2}\lsim 10$ GeV$^{2}$, our calculations
show  a weak sensitivity of $A_{z}$ to variations of
$\alpha_{iN}$ for the diagonal $iN\rightarrow iN$ rescatterings
of the excited states $i$. However, the effect of the
uncertainty associated with the
$\alpha(iN\rightarrow kN)$'s is not
negligible. The hatched area In Fig.~2 shows the
band of variation  of $A_{z}$ corresponding to
variations of $\alpha(iN\rightarrow kN)$
between -0.5 and 0.5.
 It appears that, at least for
nuclear mass number $A \sim 10$, the uncertainty of the
theoretical predictions does not rule out
 the possibility of using the $Q^{2}$ dependence of
$A_{z}$ in the  $Q^{2}$ range attainable at CEBAF to explore
the onset of CT.

In conclusion, we have performed a MST calculation of the
longitudinal asymmetry $A_{z}$ in $(e,e'p)$ scattering. Our analysis
improves upon
the previous works on this problem in several aspects.
We have for the first time taken into account
the background asymmetry induced by the nonvanishing ratio
between the real and the imaginary part of the
proton- and resonance-nucleon amplitudes, which turns out to be
very important and dominates the asymmetry at moderate $Q^{2}$.
 Moreover, we have developed a
 consistent treatment of the ejectile state, which
 is
also free of problems of double counting. The onset of CT,
 connected with the occurrence
 of inelastic rescatterings, has been studied
using a realistic model of the diffraction scattering matrix, without
making approximations in the treatment of coherency effects in the
coupled channel multiple scattering series. Our results suggest
that the $Q^{2}$ dependence of $A_{z}$, if measured in
high statistics experiments off light targets, can be used as a signature
of the onset of CT in the kinematical range accessible at CEBAF.

 This work was partly supported by the
 Grant N9S000 from the International Science Foundation and
 the INTAS grant 93-239.
AAU acknowledges Prof. A. Zichichi and ICSC- World Laboratory
for financial
support.
BGZ wishes to gratefully acknowledge the hospitatlity of the
Interdisciplinary Laboratory of SISSA.

%********************************************************************
\pagebreak

\pagebreak

%******************************************************
{\large \bf Figure captions:}
%------------------------------------------------------------------
\begin{itemize}

\item[Fig.~1]

{}~- The typical diagrams contributing to the
FSI factor
$\Phi(\vec{r}_{1},\vec{r}_{1}^{\,'})$ in the MST:
(a) the diagram without the interaction between the two
trajectories outgoing from $\vec{r}_{1}$ and $\vec{r}_{1}^{\,'}$,
(b) the diagram containing the interaction between the trajectories
generated by the term
$\hat\Gamma(b_{j})\hat\Gamma^{*}(b_{j}^{'})$.

\item[Fig.~2]

{}~- The $Q^{2}$-dependence of the longitudinal
asymmetry $A_{z}(x,y)$ as given by Eq.~(1) in the case of
parallel kinematics $p_{\perp}=0$ for
$^{16}O(e,e'p)$ and $^{40}Ca(e,e'p)$ scattering.
The solid lines show the results obtained with
inclusion of the elastic and inelastic intermediate
states within the coupled-channel MST, while the dashed lines were obtained
in the one-channel Glauber model.
The shaded areas
show the
limits of variations of $A_{z}^{MST}$ for
variations of $\alpha(iN\rightarrow kN)$
between -0.5 and 0.5.
\end{itemize}

\end{document}